\documentclass[aps,showpacs,preprintnumbers,nofootinbibt, twocolumn]{revtex4}
\usepackage{graphicx}

\begin{document}

\title{Cosmological dynamics of dark matter Bose-Einstein Condensation}
\author{T. Harko}
\email{harko@hkucc.hku.hk}
\affiliation{Department of Physics and
Center for Theoretical and Computational Physics, The University
of Hong Kong, Pok Fu Lam Road, Hong Kong, P. R. China}

\begin{abstract}
Once the critical temperature  of a cosmological boson gas is less than the critical temperature,  a Bose-Einstein Condensation process can always take place during the cosmic history of the universe. In the Bose-Einstein Condensation model, dark matter can be
described as a non-relativistic, Newtonian gravitational condensate, whose density and
pressure are related by a barotropic equation of state, with barotropic index equal to one. In the present work, we study
the Bose-Einstein Condensation process in a cosmological context, by assuming that this process can be described (at least approximately) as a first order phase transition.   We
analyze the evolution of the physical quantities relevant for the
physical description of the early universe, namely, the energy
density, temperature and scale factor, before, during and after the
 Bose-Einstein Condensation (phase transition). We also consider in detail the epoch when
the universe evolved through a mixed condensate - normal dark matter phase, with  a monotonically growing Bose-Einstein dark matter component. An important parameter characterizing the Bose-Einstein Condensation is the
condensate dark matter fraction, whose time evolution describes the time dynamics of the conversion
process. The behavior of this parameter during the cosmological condensation process is also analyzed in detail. To study the cosmological dynamics and evolution
we use both analytical and numerical methods. The
presence of the condensate dark matter and of the Bose-Einstein phase transition could have modified drastically the cosmological evolution of the early universe, as well as the large scale structure formation process.
\end{abstract}


\date{\today}

\maketitle

\section{Introduction}

At very low temperatures, particles in a dilute Bose gas can
occupy the same quantum ground state, forming a Bose-Einstein
(BEC) condensate, which appears as a sharp peak over a broader
distribution in both coordinates and momentum space. The
possibility to obtain quantum degenerate gases by a combination of
laser and evaporative cooling techniques has opened several new
lines of research, at the border of atomic, statistical and
condensed matter physics (for recent reviews on the Bose-Einstein Condensation see \cite{Da99,rev, Pet}).

An ideal system for the experimental observation of the Bose-Einstein
Condensation is a dilute atomic Bose gas confined in a trap and
cooled to very low temperatures. BEC were first observed in 1995
in dilute alkali gases such as vapors of rubidium and sodium \cite{exp}. In
these experiments, atoms were confined in magnetic traps,
evaporatively cooled down to a fraction of a microkelvin, left to
expand by switching off the magnetic trap, and subsequently imaged
with optical methods. A sharp peak in the velocity distribution
was observed below a critical temperature, indicating that
condensation has occurred, with the alkali atoms condensed in the
same ground state. Under the typical confining conditions of
experimental settings, BEC's are inhomogeneous, and hence
condensates arise as a narrow peak not only in the momentum space
but also in the coordinate space \cite{exp}.

Since in the terrestrial experiments Bose-Einstein Condensation is a well-known phenomenon, the possibility that a similar condensation may have occurred during the cosmological evolution of the universe cannot be excluded {\it a priori}. In fact, once the critical temperature  of the boson gas is less than the critical temperature,  BEC can always take place at some moment during the cosmic history of the universe.  Different aspects of the BEC cosmological transition were considered in \cite{Fuk05}-{\cite{Fuk09}. The
critical temperature for the condensation to take place is $T_{cr}<2\pi \hbar
^{2}n^{2/3}/mk_{B}$, where $n$ is the particle number density, $m$ is the particle mass, and $k_B$ is Boltzmann's constant \cite{Da99}-\cite{Pet}. On the other hand, cosmic evolution has the same temperature dependence, since
in an adiabatic expansion process the density of a matter dominated universe evolves as $\rho \propto T^{3/2}$ \cite{Fuk09}. Therefore, if the boson temperature is equal, for example, to the radiation temperature at a redshift $z = 1000$,  the critical temperature for the Bose-Einstein Condensation is at present $T_{cr} = 0.0027K$ \cite{Fuk09}. Since the matter temperature $T_m$ varies as $T_m\propto  a^{-2}$, where $a$ is the scale factor of the universe, it follows that during an adiabatic evolution the ratio of the photon temperature $T_{\gamma }$ and of the matter temperature evolves as $T_{\gamma }/T_m\propto a$. Using for the present day energy density of the universe the value $\rho _{cr}= 9.44 \times 10^{-30}$ g/cm$^3$, BEC takes place provided that the boson mass satisfies the restriction $m < 1.87$ eV \cite{Fuk08}.
On the other hand,  we expect that the universe is always under critical temperature, if it is at the present time \cite{Fuk09}.

Despite the important achievements of the  $\Lambda $ Cold Dark Matter ($\Lambda $CDM)  model \cite{ac},
at galactic scales $\sim 10$ kpc, the standard cosmological model meets with
severe difficulties in explaining the observed
distribution of the invisible matter around the luminous one. In fact, $N$%
-body simulations, performed in this scenario, predict that bound halos
surrounding galaxies must have very characteristic density profiles, called the Navarro-Frenk-White (NFW) profiles, which feature a well pronounced central cusp 
$\rho _{NFW}(r)=\rho_{s}/(r/r_{s})(1+r/r_{s})^{2} $,
where $r_{s}$ is a scale radius
and $\rho _{s}$ is a characteristic density \cite{nfw}. On the observational side,
high-resolution rotation curves show, instead, that the actual distribution
of dark matter is much shallower than the above, and it presents a nearly
constant density core \cite{bur}. The core-cusp problem of standard CDM models can be solved by assuming that  dark matter is composed of ultralight scalar particles, with masses of the order of $10^{-22}$ eV, initially in a cold Bose-Einstein condensate ( "fuzzy dark matter") \cite{Hu}. The wave properties of the dark matter stabilize gravitational collapse, providing halo cores. On the other hand dark matter models with pressure, satisfying a polytropic equation of state, give an excellent fit to the observed galactic rotational curves \cite{sax}. The polytropic equation of state can describe extended theories of dark matter involving self-interaction, non-extensive thermostatistics or boson condensation (in a classical limit). In such models, the flat-cored mass profiles widely observed in disc galaxies are due to innate dark physics, regardless of any baryonic motion.  The fine-structure in the observed inner mass distribution of the Milky Way can be explained only if
the infalling dark matter particles, from which such galactic halos formed, had
a net overall rotation, causing a ¡¦tricusp¡¦ caustic ring of dark matter \cite{duffy}. For standard, non-interacting CDM models, however, one expects the infall to be irrotational.  Bose-Einstein condensate dark matter can
form vortices, thus leading to a net overall rotation of the galactic halos. Hence
Milky-Way observations may have already detected some specific signatures of the Bose-Einstein Condensate dark matter \cite{sik}.

The possibility that the galactic dark matter is in the form of a (cold) Bose-Einstein
condensate  was considered in detail in \cite{BoHa07}. The density
distribution $\rho $ of the static gravitationally bounded single
component dark matter Bose-Einstein condensate is given by  $\rho \left( r\right) =\rho _{c}\sin kr/kr$,
where  $\rho _{c}$ is the central
density of the condensate, $\rho _{c}=\rho (0)$, and $k$ is a constant.  At the boundary of the dark matter
distribution $\rho (R)=0$, giving the condition $kR=\pi $,
which fixes the radius of the condensate dark matter halo as $R=\pi
\sqrt{\hbar ^{2}l_a/Gm^{3}}$, where $l_a$ is the $s$-wave scattering length, and $m$ is the particle mass.
The total mass of the condensate dark matter halo $M$ can be obtained as
$M=4\pi ^2\left(\hbar ^2l_a/Gm^3\right)^{3/2}\rho _c=4R^3\rho _c/\pi $,
giving for the mean value $<\rho >$ of the condensate density the expression $<\rho >=3\rho _c/\pi ^2$. The mass of the particle in the condensate is given by
\begin{eqnarray}\label{becmass}
 m &=&\left( \frac{\pi ^{2}\hbar ^{2}l_a}{GR^{2}}\right) ^{1/3}\approx
6.73\times 10^{-2}\times \nonumber\\
&&\left[ l_a\left( {\rm fm}\right) \right] ^{1/3}%
\left[ R\;{\rm (kpc)}\right] ^{-2/3}\;{\rm eV}.
\end{eqnarray}

The global cosmological evolution and the evolution of the density contrast in the Bose-Einstein condensate dark matter model, in the framework of a post-Newtonian cosmological approach, were investigated in \cite{Har} and \cite{Chav1}.  The global cosmological evolution as well as the evolution of the perturbations of the condensate dark matter shows significant differences with respect to the pressureless
dark matter model, considered in the framework of standard cosmology. Therefore the
presence of condensate dark matter could have modified drastically the cosmological
evolution of the early universe, as well as the large scale structure formation process. The core/cusp problem and the dark halo properties of the dwarf galaxies were recently analyzed in \cite{Har1}. Different properties of the dark matter condensate have been extensively studied in the physical literature \cite{lit}.

It is the purpose of the present paper to investigate the cosmological Bose-Einstein Condensation process of dark matter. The condensation process is interpreted as a  phase transition taking place sometimes during the cosmic history of the universe. A Bose-Einstein Condensation process of the dark matter in the expanding
universe can be described generically as follows.
As the normal bosonic dark matter
cools below the critical temperature $T_{cr}$ of the condensation, it
becomes energetically favorable to form a condensate, in which all particles are in the same quantum state. However, the new phase does not appear instantaneously, and the two phases coexist for some time.  The transition ends when all normal dark matter has been converted to a condensed state. During the transition phase the global cosmological evolution of the universe is changed.  By assuming that the phase transition is of
the first order, we study in detail the evolution of the relevant
cosmological parameters (energy density, temperature, scale
factor, etc) of the normal dark matter and Bose-Einstein condensed dark matter phase, and the condensation process
 itself.  An important parameter to describe the Bose-Einstein Condensation is the
condensate dark matter fraction, whose time evolution describes the time dynamics of the conversion
process. The behavior of this parameter is also analyzed in detail.

The present paper is organized as follows. In Section
\ref{sect2}, we briefly outline, for self-completeness and
self-consistency, the basic properties of the normal and Bose-Einstein condensed dark matter. We also
lay down the equations of state and the relevant physical
quantities that are analyzed in the remaining Sections. The cosmological dynamics of the Bose-Einstein Condensation of dark matter is analyzed in Section~\ref{sect3}. We discuss and conclude our results in Section~\ref{sect4}. In the present paper we use the CGS system of units.

\section{Cosmological dark matter in normal and Bose-Einstein condensed states}\label{sect2}

In the present Section we outline the relevant physical processes and quantities  of
the Bose-Einstein Condensation, which will be used in the
following Sections to study the dynamics of the condensation and some of its cosmological implications.

We assume that the space-time geometry is the flat Friedmann-Robertson-walker (FRW) metric, given by
\begin{equation}
ds^{2}=-c^2dt^{2}+a^{2}(t)\left( dx^{2}+dy^{2}+dz^{2}\right),
\label{R6}
\end{equation}
where $a$ is the scale factor describing the cosmological expansion. For the matter energy-momentum tensor on the we restrict our analysis to the case of the perfect fluid energy-momentum tensor,
\begin{equation}
T^{\mu \nu }=(\rho c^2+p)u^{\mu }u^{\nu }+pg^{\mu \nu }.
\end{equation}

The Hubble function $H(a)$ is defined as $H=\dot{a}/a$. As for the matter content of the universe, we assume that it consists of radiation, with energy density $\rho _{rad}$ and pressure $p_{rad}$, pressureless ($p_b=0$) baryonic matter, with energy density $\rho _b$, and dark matter, with energy density $\rho _{\chi}$, and pressure $p_{\chi }$, respectively. In the following we neglect any possible interaction between these components, by assuming that the energy of each component is individually conserved. Thus, the gravitational field equations, corresponding to the line element (\ref{R6}) become
\begin{equation}
3\frac{\dot{a}^{2}}{a^{2}} =8\pi G\left(\rho _b+\rho _{rad}+\rho _{\chi }\right) +\Lambda ,  \label{dH}
\end{equation}
\begin{equation}
2\frac{\ddot{a}}{a}+\frac{\dot{a}^{2}}{a^{2}} = -\frac{8\pi G}{c^2}\left(p_b+p_{rad}+p_{\chi}\right)+\Lambda,  \label{dVHi}
\end{equation}
\begin{equation}
\dot{\rho _i}+3\left(\rho _i +\frac{p_i}{c^2}\right)\frac{\dot{a}}{a} =0,\;  i=b, rad, \chi . \label{drho}
\end{equation}

The cosmological evolution of the energy density of the baryonic matter and radiation are given by $\rho _b=\rho _{b,0}/\left(a/a_0\right)^3$ and $\rho _{rad}=\rho _{rad,0}/\left(a/a_0\right)^4$, respectively, where $\rho _{b,0}$ and $\rho _{rad,0}$ are the energy densities of the matter corresponding to the value $a=a_0$ of the scale factor. For the dark matter we consider a general density evolution of the form $\rho _{\chi}=\rho _{\chi ,0}/f\left(a/a_0\right)$, where $\rho _{\chi ,0}$ is the value of the energy density of the dark matter at $a=a_0$, and $f\left(a/a_0\right)$ is an arbitrary function of the scale factor, depending on the particular dark matter model. By introducing the critical density at $a=a_0$ as $\rho _{cr,0}=3H_0^2/8\pi G$, where $H_0=H\left(a_0\right)$, and the density parameters $\Omega _{i,0}=\rho _{i,0}/\rho _{cr,0}$, $i=b,rad,\chi $, we obtain the basic equation describing the dynamics of the cosmological models as
\begin{equation}
\frac{\dot{a}^{2}}{a^{2}}=H_0^2\left[\frac{\Omega _{b,0}}{\left(a/a_0\right)^3}+\frac{\Omega _{rad,0}}{\left(a/a_0\right)^4}+\frac{\Omega _{\chi,0}}{f\left(a/a_0\right)}+\Omega _{\Lambda }\right],
\end{equation}
where $\Omega _{\Lambda }$ is the density parameter of the dark energy. The density parameters satisfy the relation $\Omega _{b,0}+\Omega _{rad,0}+\Omega _{\chi,0}+\Omega _{\Lambda }=1$.

\subsection{Normal dark matter in the early universe}

We assume that in the early stages of the evolution of the universe dark matter consisted of bosonic particles of mass $m_{\chi}$ and temperature $T$, originating in equilibrium and decoupling at a temperature $T_D$ or chemical potential $\mu >>m_{\chi }$. By assuming that the dark matter forms an isotropic gas of particles in kinetic equilibrium, the spatial number density is given by
\begin{equation}
n=\frac{g}{h^3}\int{4\pi f(p)p^2dp},
\end{equation}
where $h$ is Planck's constant, $g$ is the number of helicity states, and
\begin{equation}
f(p)=\left\{\exp\left[\left(E-\mu \right)\right]-1\right\}^{-1},
\end{equation}
where $p$ is the momentum of the particle and $E=\sqrt{p^2+m_{\chi }^2c^4}$ is the energy. A particle species that decouples in the early universe from the remaining plasma at temperature $T_D$ redshifts its momenta according to $p(t)=p_Da_D/a(t)$, where $a(t)$ is the cosmological scale factor and $a_D$ is the value of the scale factor at the decoupling. The number density of the particles $n$ evolves as $n_{\chi }\sim a^{-3}(t)$ \cite{mad}. The distribution function $f$ at a time $t$ after the decoupling is related to the value of the  distribution function at the decoupling by $f(p)=f\left(pa/a_D\right)$. The distribution function keeps an equilibrium shape in two regimes. In the extreme-relativistic case, when $E\approx pc$, $T=T_Da_D/a$, and $\mu =\mu _Da_D/a$, respectively, the distribution function is given by $f_{ER}(p)=\left\{\exp\left[\left(pc-\mu \right)\right]-1\right\}^{-1}$. In the non-relativistic decoupling case $E-\mu \approx p^2/2m_{\chi }-\mu _{kin}$, where we have defined $\mu _{kin}\equiv \mu -m_{\chi }c^2$, the distribution function is $f_{NR}(p)=\left\{\exp\left[\left(p^2/2m_{\chi }-\mu _{kin} \right)\right]-1\right\}^{-1}$. In the non-relativistic case $\mu _{kin}$ and $T$ evolve as $\mu _{kin}=\mu _{kin,D}\left(a_D/a\right)^2$ and $T=T_D/\left(a_D/a\right)^2$, respectively \cite{mad}.

The kinetic energy-momentum tensor $T^{\mu }_{\nu }$ associated to the frozen distribution of dark matter is given by
\begin{equation}
T^{\mu }_{\nu }=\frac{g}{h^3}\int{\frac{p^{\mu }p_{\nu }}{p^0}f(p)d^3p}.
\end{equation}
The energy density $\epsilon $ of the system is defined as
\begin{equation}
\epsilon =\frac{g}{3h^3}\int{Ef(p)d^3p},
\end{equation}
while the pressure of a system with an isotropic distribution of momenta is given by
\begin{equation}
P=\frac{g}{3h^3}\int{pvf(p)d^3p}=\frac{g}{3h^3}\int{\frac{c^2p^2}{E}f(p)d^3p},
\end{equation}
where the velocity $v$ is related to the momentum by $v=pc^2/E$ \cite{shap}.
In the non-relativistic regime, when $E\approx m_{\chi }c^2$ and $p\approx m_{\chi}v_{\chi }$,   the density $\rho _{\chi }$ of the dark matter is given by $\rho _{\chi }=m_{\chi }n_{\chi }$, while its pressure $P_{\chi }$ can be obtained as \cite{mad}
\begin{equation}
P_{\chi}=\frac{g}{3h^3}\int{\frac{p^2c^2}{E}f(p)d^3p}\approx 4\pi \frac{g}{3h^3}\int{\frac{p^4}{m_{\chi }}dp},
\end{equation}
giving
\begin{equation}\label{pres1}
P_{\chi}=\rho _{\chi }c^2\sigma ^2,
\end{equation}
where $\sigma ^2=\langle \vec{v}^{\;2}_{\chi } \rangle /3c^2$, and $\langle \vec{v}^{\;2}_{\chi } \rangle$ is the average squared velocity of the particle. $\sigma $ is the one-dimensional velocity dispersion. The cosmological dynamics of the dark matter density is described by the equation
\begin{equation}
\dot{\rho _{\chi}}+3\rho _{\chi}(1+\sigma ^2)\frac{\dot{a}}{a} =0,
\end{equation}
with the general solution given by
\begin{equation}
\rho _{\chi }=\frac{\rho _{\chi ,0}}{\left(a/a_0\right)^{3\left(1+\sigma ^2\right)}},
\end{equation}
where $\rho _{\chi ,0}$ is the density of the dark matter at $a=a_0$. The dynamics of the universe with normal dark matter component is described by the equation
\begin{equation}
\frac{\dot{a}^{2}}{a^{2}}=H_0^2\left[\frac{\Omega _{b,0}}{\left(a/a_0\right)^3}+\frac{\Omega _{rad,0}}{\left(a/a_0\right)^4}+\frac{\Omega _{\chi ,0}}{\left(a/a_0\right)^{3\left(1+\sigma ^2\right)}}+\Omega _{\Lambda }\right].
\end{equation}

In the following we consider that $a=a_0$ is the present day scale factor. Therefore for the Hubble constant we adopt   the value $%
H_{0}=70\;{\rm km}/{\rm s}/{\rm Mpc}=2.27\times 10^{-18}\;{\rm s}^{-1}$, giving for the critical
density a value of $\rho _{cr,0}=9.24\times 10^{-30}\;{\rm g}/{\rm cm}^{3}$. The present day dark matter density
parameter is $\Omega _{\chi ,0}\approx 0.228$, $\Omega _{b,0}=0.045$, $\Omega _{rad,0}=8.24\times 10^{-5}$, and
$\Omega _{\Lambda }=0.73$, respectively \cite{Hin09}. We also introduce the Hubble time, defined as
$t_H=1/H_0=4. 39\times 10^{17}$ s.
%The time evolution of the scale factor in the presence of normal dark matter is represented for $\sigma ^2=3.7037\times 10^{-8}$ ($\langle\vec{v}^2\rangle=10^{14}$ cm$^2$/s$^2$), in Fig~\ref{fig1}.
%\begin{figure}[!ht]
%\includegraphics[width=0.98\linewidth]{cosm1.eps}
%\caption{Time evolution of the scale factor of the universe with normal dark matter component for $\sigma ^2=3.7037\times 10^{-8}$.} \label{fig1}
%\end{figure}
Since dark matter is non-relativistic, the global cosmological evolution of the universe is very little influenced by the variation of the numerical values of $\sigma ^2$.

\subsection{Bose-Einstein condensed dark matter}\label{2}

At very low temperatures, all particles in a dilute Bose gas condense to the
same quantum ground state, forming a Bose-Einstein Condensate (BEC). Particles become correlated with each other when their wavelengths
overlap, that is, the thermal wavelength $\lambda _{T}$ is greater than the
mean inter-particles distance $l$. This happens at a temperature $T_{cr}\approx 2\pi \times
\hbar ^{2}\rho ^{2/3}/m^{5/3}k_{B}$, where $m$ is the mass of the particle in the
condensate, $\rho $ is the  density, and $k_{B}$ is Boltzmann's constant
\cite{Da99}. A coherent state develops when the particle density is enough
high, or the temperature is sufficiently low.
We assume that the dark matter halos are composed of a strongly - coupled dilute Bose-Einstein
condensate at absolute zero. Hence almost all the dark matter particles are
in the condensate.  In a dilute and cold gas,
only binary collisions at low energy are relevant, and these collisions are
characterized by a single parameter, the $s$-wave scattering length $l_a$,
independently of the details of the two-body potential. Therefore, one can
replace the interaction potential with an effective
interaction $V_I\left( \vec{r}^{\;\prime }-\vec{r}\right) =\lambda \delta \left(
\vec{r}^{\;\prime }-\vec{r}\right) $, where the coupling constant $\lambda $
is related to the scattering length $l_a$
through $\lambda =4\pi \hbar ^{2}l_a/m_{\chi }$ \cite{Da99}.
The ground state properties of the dark matter are
described by the mean-field Gross-Pitaevskii (GP) equation. The GP equation
for the dark matter halos can be derived from the GP energy functional,
\begin{eqnarray}
E\left[ \psi \right] &=&\int \left[ \frac{\hbar ^{2}}{2m_{\chi }}\left| \nabla \psi
\left( \vec{r}\right) \right| ^{2}+\frac{U_{0}}{2}\left| \psi \left( \vec{r}%
\right) \right| ^{4}\right] d\vec{r}-\nonumber\\
&&\frac{1}{2}Gm_{\chi }^{2}\int \int \frac{\left|
\psi \left( \vec{r}\right) \right| ^{2}\left| \psi \left( \vec{r}^{\;\prime
}\right) \right| ^{2}}{\left| \vec{r}-\vec{r}^{\;\prime }\right| }d\vec{r}d%
\vec{r}^{\;\prime },
\end{eqnarray}
where $\psi \left( \vec{r}\right) $ is the wave function of the condensate,
and $U_{0}=4\pi \hbar ^{2}l_a/m_{\chi }$ \cite{Da99}. The first term in the energy functional is
the quantum pressure, the second is the interaction energy, and the third is
the gravitational potential energy. The mass density of the condensate dark matter is
defined as
\begin{equation}
\rho _{\chi }\left( \vec{r}\right) =m_{\chi }\left| \psi \left( \vec{r}\right)
\right| ^{2}=m_{\chi}\rho \left(\vec{r},t\right),
\end{equation}
and the normalization condition is $N=\int \left| \psi \left(
\vec{r}\right) \right| ^{2}d\vec{r}$, where $N$ is the total number of dark
matter particles. The variational
procedure  $\delta E\left[ \psi \right] -\mu \delta \int \left| \psi \left(
\vec{r}\right) \right| ^{2}d\vec{r}=0$ gives the GP equation as
\begin{eqnarray}
&&-\frac{\hbar ^{2}}{2m_{\chi }}\nabla ^{2}\psi \left( \vec{r}\right) +m_{\chi }V\left( \vec{r}%
\right) \psi \left( \vec{r}\right) + \nonumber\\
&&U_{0}\left| \psi \left( \vec{r}\right)
\right| ^{2}\psi \left( \vec{r}\right) = \mu \psi \left( \vec{r}\right) ,
\end{eqnarray}
where $\mu $ is the chemical potential, and the gravitational potential $V$
satisfies the Poisson equation $\nabla ^{2}V=4\pi G\rho $.
In the time-dependent case the generalized Gross-Pitaevskii equation describing a
gravitationally trapped rotating Bose-Einstein condensate is given by
\begin{eqnarray}\label{sch}
i\hbar \frac{\partial }{\partial t}\psi \left( \vec{r},t\right)& =&
\left[ -%
\frac{\hbar ^{2}}{2m_{\chi }}\nabla ^{2}+m_{\chi }V\left(
\vec{r}\right) +U_0\left| \psi \left( \vec{r},t\right)
\right| ^{2} \right] \times \nonumber\\
&&\psi \left( \vec{r},t\right) .
\end{eqnarray}

The physical properties of a Bose-Einstein condensate described by the
generalized Gross-Pitaevskii equation given by Eq.~(\ref{sch}) can be
understood much easily by using the so-called Madelung representation of the
wave function \citep{Da99,rev,Pet}, which consist in writing $\psi $ in the form
\begin{equation}
\psi \left( \vec{r},t\right) =\sqrt{\rho \left( \vec{r},t\right) }\exp \left[
\frac{i}{\hbar }S\left( \vec{r},t\right) \right] ,
\end{equation}
where the function $S\left( \vec{r},t\right) $ has the dimensions of an
action. By substituting the above expression of  $\psi \left( \vec{r}%
,t\right) $ into Eq.~(\ref{sch}), it decouples into a system of two
differential equations for the real functions $\rho _{\chi }$ and $\vec{v}$,
given by
\begin{equation}
\frac{\partial \rho _{\chi }}{\partial t}+\nabla \cdot \left( \rho _{\chi }\vec{v}%
\right) =0,  \label{cont}
\end{equation}
\begin{eqnarray}
\rho _{\chi }\left[ \frac{\partial \vec{v}}{\partial t}+\left( \vec{v}\cdot
\nabla \right) \vec{v}\right] &=&-\nabla P_{\chi }\left(\frac{\rho _{\chi }}{m_{\chi }}\right)
 -\nonumber\\
 &&\rho _{\chi }\nabla \left(
\frac{V}{m_{\chi }}\right) -\nabla V_{Q},  \label{euler}
\end{eqnarray}
where we have introduced the quantum potential
$V_{Q}=-\left(\hbar ^{2}/2m_{\chi }\right)\nabla ^{2}\sqrt{\rho _{\chi }}/\sqrt{\rho
_{\chi }}$, and the velocity of the quantum fluid $\vec{v}=\nabla S/m_{\chi }$,
respectively. The effective pressure of the condensate is given by
\begin{equation}\label{pres2}
P_{\chi }\left( \frac{\rho _{\chi }}{m_{\chi }}\right) =u_{0 }\rho _{\chi }^{2},
\label{state}
\end{equation}
where
\begin{equation}
u_{0}=\frac{2\pi \hbar ^2 l_a}{m_{\chi }^3}.
\end{equation}

The Bose-Einstein gravitational condensate can be described as a gas whose
density and pressure are related by a polytropic equation of state,  with index $n=1$ \cite{BoHa07}.
When the number of particles in the gravitationally bounded Bose-Einstein
condensate becomes large enough, the quantum pressure term makes a
significant contribution only near the boundary of the condensate. Thus the quantum
stress term in the equation of motion of the condensate can be neglected.
This is the Thomas-Fermi approximation, which has been extensively used for
the study of the Bose-Einstein condensates \cite{Da99, rev, Pet}. As the number of
particles in the condensate becomes infinite, the Thomas-Fermi approximation
becomes exact. This approximation also corresponds to the
classical limit of the theory. From its definition it follows that the velocity field is irrotational,
satisfying the condition $\nabla \times \vec{v}=0$.

The cosmological evolution of the energy density of the Bose-Einstein condensate is determined by the equation
\begin{equation}
\dot{\rho }_{\chi }+3\rho _{\chi }\left(1+\frac{u_0}{c^2}\rho _{\chi }\right)\frac{\dot{a}}{a}=0,
\end{equation}
with the general solution given by
\begin{equation}
\rho _{\chi }=\frac{C_{\chi }}{\left(a/a_0\right)^3-\left(u_0/c^2\right)C_{\chi }},
\end{equation}
where $C_{\chi }$ is an arbitrary constant of integration. By using the condition that $\rho _{\chi }=\rho _{\chi ,0}$ for $a=a_0$, we obtain the density of the condensate in the form
\begin{equation}\label{29}
\rho _{\chi }=\frac{c^2}{u_0}\frac{\rho _{0\chi }}{\left(a/a_0\right)^3-\rho _{0\chi }},
\end{equation}
where we have denoted
\begin{equation}
\rho _{0\chi }=\frac{\rho _{\chi ,0}u_0/c^2}{1+\rho _{\chi ,0}u_0/c^2}= \frac{\rho _{cr,0}\Omega _{\chi, 0}u_0/c^2}{1+\rho _{cr,0}\Omega _{\chi, 0}u_0/c^2}.
\end{equation}

\section{Cosmological dynamics of the Bose-Einstein Condensation}\label{sect3}

The order of the phase transition of the Bose-Einstein Condensation
in an interacting Bose systems, i.e, the passage from the normal to the condensed phase
where all particles occupy a single-particle state, has been intensively discussed in the recent physical literature. According to some results the Bose-Einstein Condensation shows a spontaneous $U(1)$ gauge
symmetry breaking, with the condensate fraction $N_0/N$ playing the role of the order
parameter \cite{ft2}. That would suggest a second order phase transition. However, by analyzing the
dependence of the chemical potential $\mu $ on the temperature $T$ and particle density $\rho $ in the framework of several theoretical models, describing the thermodynamical transition from the normal to the
BEC phase in weakly interacting Bose gases, it was shown that none of them predicts a second-order phase transition, as required by symmetry-breaking general considerations \cite{ft}. These results would imply that the Bose-Einstein Condensation represents a first order phase transition \cite{ft,tr}. On the other hand, by definition, first-order phase transitions in thermodynamics feature a genuine mathematical singularity. Whether finite systems in nature can literally exhibit such a behavior is a long-standing controversial question in physics \cite{cont}. However, a rigorous investigation into the thermodynamic instability of an ideal Bose gas, confined in a cubic
box has shown that a system consisting of a finite number of particles can exhibit a discontinuous phase transition that features a genuine mathematical singularity, provided that not the volume,  but the pressure is kept constant \cite{tr}. This result was obtained without assuming a thermodynamic limit or a continuous approximation. Hence, it seems presently that the best possible description of the intrinsic dynamics of the Bose-Einstein Condensation can be obtained within the framework of a first order phase transition.

\subsection{Cosmological parameters at the condensation point}

Generally, the chemical potential $\mu $ of a physical system is a function of the temperature and particle density $n=N/V$,  $\mu =\mu \left(n, T\right)$, where $N$ is the total particle number, and $V$ is the volume. The extensivity property of the Helmholtz free energy $F = F(N, V, T)$ allows us to either write $F = V f(n, T)$ or
$F = N \tilde{f}(v, T)$, with $v = V/N =n^{-1}$,  such that $f = n\tilde{f}$. Hence both forms carry the same
physical information. From the Helmholtz free energy we can obtain the chemical potential as $\mu (n,T)=\left(\partial f/\partial n\right)_T$ and the pressure as $p(n,T)=-\left(\partial \tilde{f}/\partial v\right)_T$. Therefore, $\mu = \mu (n,T)$ and $p = p(v, T)$  carry the same information.  The laws of thermodynamics require
that both the chemical potential and the pressure are single valued, that is, for any given values of $n$ and $T$,
or $v$ and $T$, there must only exist a single value of $\mu $  or $p$, respectively \cite{ft}.

Therefore, a first thermodynamic condition that must be satisfied during the cosmological Bose-Einstein Condensation process is the continuity of the pressure at the transition point. With the use of Eqs.~(\ref{pres1}) and (\ref{pres2}) the continuity of the pressure uniquely fixes the critical transition density $\rho _{\chi }^{cr}$ from the normal dark matter state to the Bose-Einstein condensed state as
\begin{equation}\label{crde}
\rho _{\chi }^{cr}=\frac{c^2\sigma ^2}{u_0}=\frac{c^2\sigma ^2m_{\chi }^3}{2\pi \hbar ^2 l_a}.
\end{equation}

The numerical value of the transition density depends on three unknown parameters, the dark matter particle mass, the scattering length, and the dark matter particles velocity dispersion, respectively. By assuming a typical mass of the dark matter particle of the order of 1 eV (1 eV = $1.78\times 10^{-33}$ g), a typical scattering length of the order of $10^{-10}$ cm, and a mean velocity square of the order of $\langle\vec{v}^2\rangle=81\times 10^{14}\;{\rm cm^2/s^2}$, the critical transition density can be written as
\begin{eqnarray}
\rho _{\chi }^{cr}&=&3.868\times 10^{-21}\left(\frac{\sigma ^2}{3\times 10^{-6}}\right)\times \nonumber\\
&&\left(\frac{m_{\chi }}{10^{-33}\;{\rm g}}\right)^3\left(\frac{l_a}{10^{-10}\;{\rm cm}}\right)^{-1}\;{\rm g/cm^3}.
\end{eqnarray}

The critical temperature at the moment of Bose-Einstein condensate transition is given by \cite{Da99}-\cite{Pet}
\begin{eqnarray}
T_{cr}&\approx &\frac{2\pi \hbar ^2}{\zeta (3/2)^{2/3}m_{\chi }^{5/3}k_B}\left(\rho _{\chi }^{cr}\right)^{2/3}=\nonumber\\
&&\frac{\left(2\pi \hbar ^2\right) ^{1/3}c^{4/3}}{\zeta (3/2)^{2/3}k_B}\frac{\left(\sigma ^2\right)^{2/3}m_{\chi }^{1/3}}{l_a^{2/3}},
\end{eqnarray}
where $\zeta (3/2)$ is the Riemann zeta function, or
\begin{eqnarray}
T_{cr}&\approx &6.57\times10^3\times\left(\frac{m_{\chi }}{10^{-33}\;{\rm g}}\right)^{1/3}\times \nonumber\\
&&\left(\frac{\sigma ^2}{3\times 10^{-6}}\right)^{2/3}\left(\frac{l_a}{10^{-10}\;{\rm cm}}\right)^{-2/3}\;K.
\end{eqnarray}

The critical pressure of the dark matter fluid at the condensation moment can be obtained as
\begin{eqnarray}
P_{cr}&=&1.04\times 10^{-5}\left(\frac{\sigma ^2}{3\times 10^{-6}}\right)^2\times \nonumber\\
&&\left(\frac{m_{\chi }}{10^{-33}\;{\rm g}}\right)^3\left(\frac{l_a}{10^{-10}\;{\rm cm}}\right)^{-1}\;{\rm dyne/cm^2}.
\end{eqnarray}

The critical value $a_{cr}$ of the scale factor of the universe at the moment of the condensation can be obtained from Eq.~(\ref{crde}) as
\begin{eqnarray}
a_{cr}/a_0&=&\left(\frac{\rho _{\chi ,0}u_0}{c^2\sigma ^2}\right)^{1/3\left(1+\sigma ^2\right)}=\nonumber\\
&&\left(\frac{2\pi \hbar ^2l_a \rho _{cr ,0}\Omega _{\chi ,0}}{c^2\sigma ^2m_{\chi }^3}\right)^{1/3\left(1+\sigma ^2\right)},
\end{eqnarray}
giving for the critical redshift of the transition the expression
\begin{equation}
1+z_{cr}=\left(\frac{2\pi \hbar ^2l_a \rho _{cr ,0}\Omega _{\chi ,0}}{c^2\sigma ^2m_{\chi }^3}\right)^{-1/3\left(1+\sigma ^2\right)}.
\end{equation}

By using the adopted numerical values of the constants we obtain for the critical scale factor and for the critical redshift the values
 \begin{eqnarray}
a_{cr}/a_0&= &8.17\times10^{-4}\times\left(\frac{m_{\chi }}{10^{-33}\;{\rm g}}\right)^{-\left(1+\sigma ^2\right)}\times \nonumber\\
&&\left(\frac{\sigma ^2}{3\times 10^{-6}}\right)^{-1/3\left(1+\sigma ^2\right)}\times \nonumber\\
&&\left(\frac{l_a}{10^{-10}\;{\rm cm}}\right)^{1/3\left(1+\sigma ^2\right)},
\end{eqnarray}
and
\begin{eqnarray}
1+z_{cr}&= &1.22\times10^{3}\times\left(\frac{m_{\chi }}{10^{-33}\;{\rm g}}\right)^{\left(1+\sigma ^2\right)}\times \nonumber\\
&&\left(\frac{\sigma ^2}{3\times 10^{-6}}\right)^{1/3\left(1+\sigma ^2\right)}\times \nonumber\\
&&\left(\frac{l_a}{10^{-10}\;{\rm cm}}\right)^{-1/3\left(1+\sigma ^2\right)},
\end{eqnarray}
respectively.

\subsection{Cosmological evolution during the Bose-Einstein Condensation phase}

During a first order phase transition, the temperature and the pressure are
constants, $T=T_{cr}$ and $P=P_{cr}$, respectively. The entropy
$S=sa^{3}$ and the enthalpy $W=\left( \rho +p\right) a^{3}$ are
also conserved quantities. After the beginning of the phase transition the density of the dark matter $\rho _{\chi }\left(
t\right) $ decreases from $\rho _{\chi }^{cr}\left( T_{cr}\right) \equiv
\rho ^{nor}_{\chi}$ (when all the dark matter is in a normal, non-condensed form) to $\rho _{\chi }\left( T_{cr}\right) \equiv \rho ^{BEC}_{\chi}$, corresponding to the full conversion of dark matter into a condensed state.  It is convenient to replace $\rho _{\chi }\left( t\right) $ by
$h(t)$, the volume fraction of matter in the Bose-Einstein condensed phase, which is defined as
\begin{equation}
h(t)=\frac{\rho _{\chi }\left( t\right)-\rho ^{nor}_{\chi}}{\rho ^{BEC}_{\chi}-\rho ^{nor}_{\chi}}.
\end{equation}
Therefore the evolution of the dark matter energy density during the transition process is given by
\begin{equation}
\rho _{\chi }\left( t\right) =\rho ^{BEC}_{\chi}h(t)+\rho ^{nor}_{\chi}\left[ 1-h(t)\right]
=\rho ^{nor}_{\chi} \left[ 1+n_{\chi }h(t)\right] ,
\end{equation}
where we have denoted
\begin{equation}
n_{\chi }=\frac{\rho ^{BEC}_{\chi }-\rho ^{nor}_{\chi }}{\rho ^{nor}_{\chi}}.
\end{equation}
At the beginning of the Bose-Einstein Condensation process $h(t_{cr})=0$, where
$t_{cr}$ is the time corresponding to the beginning of the phase
transition, and $\rho _{\chi }\left( t_{cr}\right) \equiv \rho ^{nor}_{\chi }$. At the end of the condensation $h\left( t_{BEC}\right) =1$, where
$t_{BEC}$ is the time at which the phase transition ends,
corresponding to $\rho _{\chi }\left( t_{BEC}\right) \equiv \rho ^{BEC}_{\chi }$. For
$t>t_{BEC}$ the universe enters in the Bose-Einstein condensed dark matter phase.

From Eq.~(\ref{drho}) we obtain
\begin{equation}\label{r}
\frac{\dot{a}}{a}=-\frac{1}{3}\frac{\left( \rho ^{BEC}_{\chi }-\rho
^{nor}_{\chi }\right) \dot{h} }{\rho ^{nor}_{\chi }+P_{cr}/c^2+\left( \rho ^{BEC}_{\chi }-\rho
^{nor}_{\chi }\right) h}=-\frac{1}{3}\frac{r \dot{h}}{1+rh},
\end{equation}
where we have denoted
\begin{equation}\label{defr}
r= \frac{\rho ^{BEC}_{\chi }-\rho ^{nor}_{\chi }}{ \rho
^{nor}_{\chi }+P_{cr}/c^2}=\frac{n_{\chi}}{1+P_{cr}/\rho _{\chi }^{nor}c^2}.
\end{equation}
 Since $\rho ^{BEC}_{\chi }<\rho ^{nor}_{\chi }$, generally $r<0$, $r\in (-1,0)$,  and $n_{\chi }<0$, respectively. Eq.~(\ref{r}) immediately leads to the expression of the scale factor of the universe during the Bose-Einstein Condensation phase as
\begin{equation}
a(t)=a_{cr} \left[ 1+rh(t)\right] ^{-1/3},t\in\left(t_{cr},t_{BEC}\right),
\end{equation}
where we have used the initial condition $h\left(t_{cr}\right)=0$ and we have denoted $a_{cr}=a\left(t_{cr}\right)$. At the end of the phase transition the scale factor of the universe has the value
\begin{equation}
a_{BEC}=a\left(t_{BEC}\right) = a_ {cr} (1+r)^{-1/3}.
\end{equation}
The Bose-Einstein Condensation modifies the overall expansion rate of the universe. Consequently, during the phase transition, the pressureless  baryonic matter and the radiation density evolve according to
\begin{equation}
\rho _b=\frac{\rho _{b,0}}{\left(a_{cr}/a_0\right)^3}\left[1+rh(t)\right],t\in\left(t_{cr},t_{BEC}\right),
\end{equation}
and
\begin{equation}
\rho _{rad}=\frac{\rho _{rad,0}}{\left(a_{cr}/a_0\right)^4}\left[1+rh(t)\right]^{4/3},t\in\left(t_{cr},t_{BEC}\right),
\end{equation}
respectively.
The evolution of the fraction of the condensed matter $h(t)$ during the Bose-Einstein Condensation process is
described by the equation
\begin{widetext}
\begin{equation}\label{h1}
\frac{dh}{d\tau }=-3\left( \frac{1}{r}+h\right)\times
\sqrt{\frac{\Omega _{b,0}}{%
\left( a_{cr}/a_{0}\right) ^{3}}\left( 1+rh\right) +\frac{\Omega _{rad,0}}{%
\left( a_{cr}/a_{0}\right) ^{4}}\left( 1+rh\right) ^{4/3}+\Omega _{\chi
,nor}\left( 1+n_{\chi }h\right) +\Omega _{\Lambda }},
\end{equation}
\end{widetext}
where we have introduced a dimensionless time variable $\tau =H_{0}t$, and we
have denoted $\Omega _{\chi ,nor}=\rho _{\chi }^{nor}/\rho _{cr,0}$, respectively. Since $P_{cr}/\rho _{\chi }^{nor}c^2=\sigma ^2<<1$, from Eq.~(\ref{defr}) it follows that $r\approx n_{\chi }$, an approximation we  adopt in the followings.  If the energy density contribution of the radiation to the total energy density of the universe can be neglected, Eq.~(\ref{h1}) can be integrated exactly to give
\begin{eqnarray}
h(t)&=&\frac{\Omega _{\Lambda }^{2}}{r\Omega _{tr}}\times \nonumber\\
&&\left\{ \frac{1+\Omega
_{cond}\times \exp \left[ -\left( 3\Omega _{\Lambda }/t_{H}\right) \left(
t-t_{cr}\right) \right] }{1-\Omega _{cond}\times \exp \left[ -\left( 3\Omega
_{\Lambda }/t_{H}\right) \left( t-t_{cr}\right) \right] }\right\} ^{2}-\nonumber\\
&&\frac{
\Omega _{\Lambda }+\Omega _{tr}}{r\Omega _{tr}},
\end{eqnarray}

where we have denoted
\begin{equation}
\Omega _{tr}=\frac{\Omega _{b,0}}{%
\left( a_{cr}/a_{0}\right) ^{3}}+\Omega _{\chi ,nor}
\end{equation}
and
\begin{equation}
\Omega _{cond}=\frac{\sqrt{\Omega _{tr}+\Omega _{\Lambda }}-\Omega _{\Lambda
}}{\sqrt{\Omega _{tr}+\Omega _{\Lambda }}+\Omega _{\Lambda }},
\end{equation}
respectively.

The time interval necessary to convert the entire existing normal dark matter to the Bose-Einstein condensed phased is given by
\begin{widetext}
\begin{equation}
\Delta t_{cond}=t_{BEC}-t_{cr}=\frac{1}{3\Omega _{\Lambda }}\ln \frac{\left[
\sqrt{\Omega _{tr}+\Omega _{\Lambda }}-\Omega _{\Lambda }\right] \left[
\sqrt{\Omega _{tr}\left( 1+r\right) +\Omega _{\Lambda }}+\Omega _{\Lambda }%
\right] }{\left[ \sqrt{\Omega _{tr}+\Omega _{\Lambda }}+\Omega _{\Lambda }%
\right] \left[ \sqrt{\Omega _{tr}\left( 1+r\right) +\Omega _{\Lambda }}%
-\Omega _{\Lambda }\right] }\times t_{H}.
\end{equation}
\end{widetext}

If the dark energy density can also be neglected,  the volume fraction of matter $h(t)$ in the Bose-Einstein condensed phase evolves as
\begin{equation}
h(t)=\frac{1}{r}\left[ 1+\frac{2}{3}H_0\sqrt{\Omega _{tr}}\left( t-t_{cr}\right) \right]
^{-2}-\frac{1}{r}.
\end{equation}
In this approximation the conversion of the entire dark matter of the universe to the BEC phase is given by
\begin{equation}
\Delta t_{cond}=\frac{2}{3}\left[ \left( 1+r\right)^{-1/2}-1\right]\frac{t_H}{\sqrt{\Omega _{tr}}} .
\end{equation}

By assuming the standard values $l_a=10^{-10}$ cm, $m=1$ eV and $\sigma ^2=3\times 10^{-6}$ we obtain $\Omega _{tr}=5.02\times 10^{8}$, giving for the total condensation time an approximate value of
\begin{equation}
\Delta t_{cond}=1.31\times\left[ \left( 1+r\right)^{-1/2}-1\right]\times 10^{13}\;{\rm s}.
 \end{equation}

The time evolution of $h(t)$, as given by Eq.~(\ref{h1}) is represented, for several values of $r$, in Fig.~\ref{fig2}.

\begin{figure}[!ht]
\includegraphics[width=0.98\linewidth]{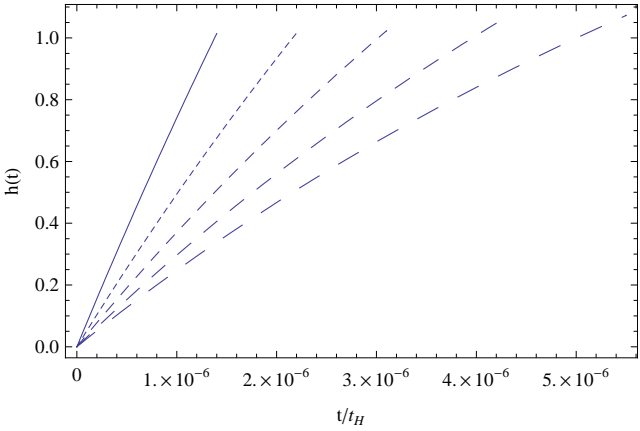}
\caption{Time evolution of the condensed dark matter fraction $h(t)$ for different values of $r$:
$r=-0.10$ (solid curve), $r=-0.15$ (dotted curve), $r=-0.20$ (dashed curve), $r=-0.25$ (long-dashed curve), and $r=-0.3$ (ultra-long dashed curve), respectively.}
\label{fig2}
\end{figure}

\subsection{The Post-Condensation phase}

At $t=t_{BEC}$ and $a_{BEC}=a_{cr}(1+r)^{-1/3}$, all the dark matter in the universe has been converted to a Bose-Einstein condensate phase. At this moment the cosmological density of the dark matter is given by
\begin{equation}
\rho _{\chi }^{BEC}=\frac{c^{2}}{u_{0}}\frac{\rho _{0\chi }\left( 1+r\right)
}{\left( a_{cr}/a_{0}\right) ^{3}-\rho _{0\chi }\left( 1+r\right) },
\end{equation}
where the constant $\rho _{0\chi }$ can be represented as
\begin{equation}
\rho _{0\chi }=\frac{1.63\times 10^{-15}\times \left(
l_{a}/10^{-10}\;{\rm cm}\right) \left( m/10^{-33}\;{\rm g}\right)^{-3} }{1+1.63\times
10^{-15}\times \left( l_{a}/10^{-10}\;{\rm cm}\right) \left( m/10^{-33}\;{\rm g}\right)^{-3} }.
\end{equation}
The condition of the positivity of the density imposes the constraint $\left(
a_{cr}/a_{0}\right) >\rho _{0\chi }^{1/3}\left( 1+r\right) ^{1/3}$ on the
model parameters.

The equation determining the time
evolution of the scale factor of the dark matter in the Bose-Einstein condensate phase is given by
\begin{equation}\label{55}
\frac{d\left( a/a_{0}\right) }{dt}=H_{0}\sqrt{\Omega _{BE}}\frac{a/a_{0}}{\sqrt{\left( a/a_{0}\right)
^{3}-\rho _{0\chi }}},
\end{equation}
where we have denoted
\begin{equation}
\Omega _{BE}=\frac{\Omega _{\chi ,0}}{%
1+\rho _{\chi ,0}u_{0}/c^{2}},
\end{equation}
and can be integrated exactly to give
\begin{eqnarray}
\sqrt{\Omega _{BE}}H_0\left(t-C\right)&=&\frac{2}{3}\sqrt{\left(\frac{a}{a_0}\right)^3-\rho _{0\chi }}-\frac{2}{3}\sqrt{\rho _{0\chi }}\times \nonumber\\
&&\arctan \left[\sqrt{\frac{\left(a/a_0\right)^3-\rho _{0 \chi }}{\rho _{0\chi }}}\right],
\end{eqnarray}
where $C$ is an arbitrary constant of integration, which can be determined from the condition $a=a_{BEC}$ at $t=t_{BEC}$. Thus we obtain
\begin{eqnarray}\label{factor}
C&=&t_{BEC}-\frac{2}{3\sqrt{\Omega _{BE}}H_0}\sqrt{\left(\frac{a_{BEC}}{a_0}\right)^3+\rho _{0\chi }}-\frac{2}{3}\times \nonumber\\
&&\sqrt{\frac{\rho _{0\chi }}{\Omega _{BE}H_0^2}}\arctan \left[\sqrt{\frac{\left(a_{BEC}/a_0\right)^3-\rho _{0 \chi }}{\rho _{0\chi }}}\right].
\end{eqnarray}

In the case of a universe filled with dark energy, radiation,
baryonic matter with negligible pressure, and Bose-Einstein condensed dark matter, respectively, the time evolution
of the scale factor is given by the differential equation
\begin{widetext}
\begin{equation}\label{59}
\frac{1}{\left( a/a_{0}\right) }\frac{d\left( a/a_{0}\right) }{dt}=H_{0}%
\sqrt{\frac{\Omega _{b,0}}{\left( a/a_{0}\right) ^{3}}+\frac{\Omega _{rad,0}%
}{\left( a/a_{0}\right) ^{4}}+\frac{\Omega _{BE}}{\left( a/a_{0}\right) ^{3}-\rho _{0\chi }}%
+\Omega _{\Lambda }},t\geq t_{BEC},
\end{equation}
\end{widetext}
which must be integrated with the initial condition $a\left(
t_{BEC}\right) =a(0)=a_{BEC}$. The time evolutions of the scale factors for universes
containing BEC dark matter  are represented, for different values of the BEC parameter
$\rho _{0\chi }$, in Fig.~\ref{fig3}.

\begin{figure}[!ht]
\includegraphics[width=0.98\linewidth]{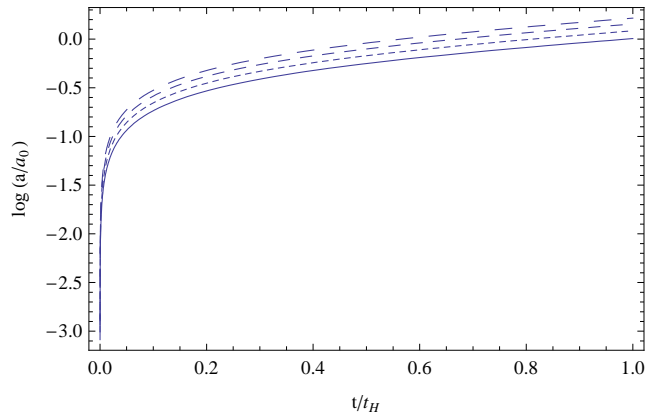}
\caption{Time evolution (in a logarithmic scale) of the scale factor of the universe in the Post-Bose-Einstein Condensation phase,
for $a_{BEC}=a(0)=8.1\times 10^{-4}$ ($z\approx 1200$), and for different values of $\rho _{0\chi }$: $\rho _{0\chi }=10^{-11}$ (solid curve), $\rho _{0\chi }=5\times 10^{-11}$ (dotted curve), $\rho _{0\chi }=10^{-10}$ (dashed curve) and $\rho _{0\chi }=5\times 10^{-10}$  (long dashed curve), respectively.}
\label{fig3}
\end{figure}

The presence of the condensate dark
matter changes the global cosmological dynamics of the universe in the post-condensation phase,
and the magnitude of the changes increases with the
increase of the BEC parameter $\rho _{0\chi }$.

\section{Discussions and final remarks}\label{sect4}

The Bose-Einstein Condensation process has been extensively studied (especially for the case of scalar fields) in the physical literature. However, many important questions have yet to be answered. For example, one would like to know more exactly if the condensation process of the dark matter takes place (almost) instantaneously, or it evolves during a long time interval. Many details of the thermodynamics of the Bose-Einstein Condensation are not yet conclusively understood. Even the order of the phase transition (crossover?) is still a matter of
debate. And, of course, from a cosmological point of view, one would like to know more exactly in what cosmological epoch the condensation did occur, what where the values of the cosmological parameters at that time, how long the transition lasted, and what were its implications on the global evolution of the universe. In the present paper we have tried to give some preliminary and qualitative answers to these questions.

Once the Bose-Einstein Condensation occurs, dark matter becomes a mixture of two phases, the normal and the condensed phase, respectively. From a general thermodynamic point of view the coexistence of these two phases requires the continuity of the pressure of the two phases at the beginning of the condensation (phase transition). This thermodynamical condition uniquely fixes the value of the dark matter density at the condensation moment, as well as the values of all the other thermodynamical parameters (temperature and pressure). The concrete numerical values of the condensation parameters depend on the mean square velocity of the normal dark matter $\sigma ^2$, and of the scattering length $l_a$ and mass $m_{\chi }$ of the dark matter particle. Since the values of $\sigma ^2$, $l_a$ and $m_{\chi }$ are not known for the dark matter particles, or are very uncertain (like $m_{\chi }$ and $\sigma $, for example), it is difficult to predict the exact cosmological moment of the Bose-Einstein condensation, and the corresponding values of the cosmological parameters. However, by adopting some "standard" numerical values, one can obtain a qualitative picture of the transition. Thus, the mass of the dark matter particle was assumed to be of the order of 1 eV \cite{Fuk09,Bo}, while the mean velocity of the non-relativistic dark matter particles was taken to be of the order of 900 km/s, which is somewhat larger than the values inferred from the study of the galactic dark matter halos \cite{Bo, mad}. The numerical value of the scattering length $l_a$ for dark matter particles is very uncertain. As can be inferred from present experiments in ultracold gases, the scattering length is a quantity that determines the thermodynamic state of the gas. Terrestrial experiments on Bose-Einstein atomic condensates give a value of $10^{-7}$ cm \cite{Da99, rev, Pet}. For simplicity we have adopted for this parameter a value of $l_a=10^{-10}$ cm.

The general analysis of the condensation process shows a distinct three-phase cosmological history of the universe. In the first phase, the universe evolved according to the standard $\Lambda $ Cold-Dark Matter ($\Lambda $CDM) model. The condensation process of the dark matter did start when the temperature of the bosonic gas did fall below the critical value, and the equality of the pressure allowed the two phases to coexist simultaneously. With the adopted numerical values of the dark matter parameters the condensation began at a redshift of around $z=1200$. The condensation process took place gradually, the normal and condensed phases coexisting for around $10^6$ years. To describe this transition period we have introduced the fraction $h(t)$ of the condensed dark matter, and studied its time evolution.  After the transition phase the universe entered in the condensed dark matter phase. Depending on the moment of its occurrence in the cosmological history,  the presence of the condensate dark matter and of the Bose-Einstein phase transition could have modified drastically the cosmological evolution of the early universe, as well as the large scale structure formation process. Of course, even small changes in the numerical values of the parameter set $\left(\sigma ^2, l_a,m_{\chi }\right)$ could lead to significant changes in the values of the critical redshift or duration of the condensation phase.
%The so-called Cardassian cosmological models were introduced in \cite{Fr} as a possible alternative to explain the acceleration of the universe by a model that has no energy components in addition to ordinary matter. These models were intensively investigated in the recent physical literature \cite{card}. In order to explain the late-time acceleration of the universe the Friedmann equation must be modified to $H^2=g\left(\rho _b\right)/3=\rho _b/3+f\left(\rho _b\right)$, where $g\left(\rho _b\right)$ and $f\left(\rho _b\right)$ are some arbitrary functions of the baryonic matter density $\rho _b$. The energy density of the pressureless baryonic matter  can be written as $\rho _b=\rho _{b,0}/\left(a/a_0\right)^3$. Then, from Eq.~(\ref{29}) it follows that the energy density of the Bose-Einstein condensate dark matter can be written as
%\begin{equation}
%\rho ^{BEC}_{\chi }=\frac{c^2\rho _{0\chi}}{u_0\rho _{b,0}}\frac{\rho _b}{1-\left(\rho _{0\chi}/\rho _{b,0}\right)\rho _b}\approx \alpha _1 \rho _b+\alpha _2 \rho _b^2,
%\end{equation}
%where $\alpha _1=c^2\rho _{0\chi }/u_0\rho _{b,0}$ and $\alpha _2=c^2\rho _{0\chi }^2/u_0\rho _{b,0}^2$, respectively. Therefore Bose-Einstein condensate dark matter models naturally belong to the Cardassian class of polytropic cosmological models.

Immediately after the phase transition the presence of a Bose-Einstein condensate may drastically change the cosmological dynamics of the universe. As one can see from Eq.~(\ref{59}), if for some time interval $\left(a/a_0\right)$ is very close to $\rho _{0\chi }^{1/3}$, the cosmological dynamic of the universe is determined by the condensed dark matter. In this case the condensed dark matter energy density $\rho ^{BEC}_{\chi }=\Omega _{BE}/\left[\left(a/a_0\right)^3-\rho _{0\chi }\right]$ is very large, and dominates all the other cosmological energy terms. When this condition is fulfilled, the expansion of the universe enters in an accelerated phase. For the time interval for which $\rho ^{BE}_{\chi }$ can be approximated as a constant, $\rho ^{BE}_{\chi }\approx$ constant, the expansion of the universe can be described as a de Sitter one. The presence of an accelerating expansion period after the phase transition can also be seen from the analysis of the deceleration parameter $q=-a\ddot{a}/\dot{a}^2=d(1/H)/dt-1$ of the universe. For $\left(a/a_0\right)\rightarrow\rho _{0\chi }^{1/3}$, the scale factor of the universe with Bose-Einstein condensate dark matter, given by Eq.~(\ref{factor}), can be approximated as
\begin{equation}
\frac{a}{a_0}\approx\left[\frac{9}{4}\Omega _{BE}H_0^2\left(t-C\right)^2+\rho _{0\chi }\right]^{1/3}.
\end{equation}
During this phase the deceleration parameter $q$ is given by
\begin{equation}
q\approx\frac{1}{2}-\frac{2\rho _{0\chi}}{3\Omega _{BE}H_0^2(t-C)^2}.
\end{equation}
During the time interval given by $t<C+\sqrt{(4/3)\rho _{0\chi}/\Omega _{BE}H_0^2}$ the deceleration parameter satisfies the condition $q<0$. However, in the limit of large times, the expansion of the universe decelerates, with the deceleration parameter given by $q\approx 1/2$.

Therefore, the Bose-Einstein condensation of the dark matter may also provide the explanation for the observed recent acceleration of the universe,  which could be just a transient phenomenon related to the recent entering of the universe in the Post-Bose-Einstein condensation phase. Hence BEC could also provide an alternative, physically realistic solution to the dark energy problem (for reviews of the dark energy problem see \cite{ac}). However, such a model would require that the cosmological BEC phase transition took place very recently, at $z\approx 1-2$, on a global cosmic scale. Presently there is no astrophysical evidence for such a scenario.

A central problem in the theory of the Bose-Einstein condensation is the nature of the dark matter particle, as well as its physical properties.  The core/cusp problem can be solved  if the dark matter is
composed of ultralight scalar particles with mass $m= 10^{-22}$ eV, initially in a (cold) Bose-Einstein condensate \cite{Hu}. The wave properties of the dark matter stabilize gravitational
collapse, providing halo cores and sharply suppressing small-scale linear power. Ultra-light dark matter particles with masses of the order of $m\sim 10^{-23}$ eV and with Compton wavelengths of the order of galactic scales have temperatures of about 0.9 K, a temperature substantially lower than the temperature of CMB neutrinos. Therefore such a mass would guarantee that Big Bang Nucleosynthesis remains unaffected. In addition, the temperature is consistent with WMAP observations without fine-tuning \cite{bond}. It was shown that axions with mass $m\approx 6\times 10^{-6}\;{\rm eV}\;10^{12}\;{\rm GeV}/f$ also form a BEC, where $f\gtrsim 10^9\;{\rm GeV}$ \cite{sik}. This would imply an axion mass of the order of $m\approx 10^{-3}$ eV.  The axions were postulated shortly after the standard model of elementary particles was established to explain why the strong interactions conserve the discrete symmetries P and CP (for more details see \cite{sik} and references therein). 
A cosmological bound on the mass of the condensate particle can be obtained as $m<2.696\left(g_d/g\right)\left(T_d/T_{cr}\right)^3$ eV \cite{Bo}, where $g$ is the number of internal degrees of freedom of the particle before decoupling, $g_d$ is the number of internal degrees of freedom of the particle at the decoupling, and $T_d$ is the decoupling temperature. In the Bose condensed case $T_d/T_c < 1$, and it follows that the BEC particle should be light, unless it decouples very early on, at high temperature and with a large $g_d$. Therefore,  depending on the relation between the critical  and the decoupling temperatures, in order for a BEC light relic to act as cold dark matter,  the decoupling scale must be higher than the electroweak scale \cite{Bo}.
On the other hand, from Eq.~(\ref{becmass}) for $l_a\approx 1 $ fm and $R\approx 10$ kpc, it follows that the typical mass of the condensate particle is of the order of $m\approx 14$ meV. For $l_a\approx 10^{6}$ fm, corresponding
to the values of $a$ observed in terrestrial laboratory experiments, $m\approx 1.44$ eV. 
These values are consistent with the limit $m<1.87$ eV obtained for the mass of the condensate particle from
cosmological considerations \cite{Fuk08}.

A better understanding of the numerical values of the Bose-Einstein condensation parameters (scattering length and dark matter particle mass), would be very helpful in obtaining accurate cosmological conclusions in the framework of the BEC model. Such an advance may also provide a powerful method for observationally testing on a cosmological scale the theoretical predictions of the Bose-Einstein Condensation model,  and the possible
existence of the condensed dark matter.

\acknowledgments
 This work is supported by a GRF grant of the government of the Hong Kong SAR.

\end{document}